\documentclass[prl,twocolumn,showpacs,flushleft,amsmath,amssymb,nofootinbib]{revtex4}
\usepackage{subfig}
\usepackage{graphicx}
\usepackage{dcolumn} 
\usepackage{bm}      
\usepackage[active]{srcltx} 
\usepackage{hyperref} 
\usepackage{amsmath}
\usepackage{amssymb}
\usepackage{stmaryrd}
\usepackage{amsfonts}
\usepackage{mathrsfs}
\usepackage{txfonts}

\newcommand{\Ref}[1]{(\ref{#1})}

\newcommand{\Z}{\mathbb{Z}}
\newcommand{\R}{\mathbb{R}}
\newcommand{\C}{\mathbb{C}}

\newcommand{\half}{\frac{1}{2}}

\newcommand{\Slc}{\mathrm{SL}(2,\mathbb{C})}

\def\be{\begin{eqnarray}}
\def\ee{\end{eqnarray}}

\newcommand{\ch}{\mathcal H}

\newcommand{\ck}{\mathcal K}

\newcommand{\cm}{\mathcal M}

\newcommand{\calr}{\mathcal R}
\newcommand{\cs}{\mathcal S}

\newcommand{\cv}{\mathcal V}

  \newcommand{\Fl}{\mathfrak{L}}
  
  \newcommand{\Fn}{\mathfrak{N}}

\newcommand{\fs}{\mathfrak{s}}

\renewcommand{\a}{\alpha}
\renewcommand{\b}{\beta}
\newcommand{\g}{\gamma}

\newcommand{\eps}{\varepsilon}

\renewcommand{\l}{\lambda}

\renewcommand{\O}{\Omega}

\newcommand{\rmd}{\mathrm d}

\newcommand{\lt}{\left}
\newcommand{\rt}{\right}

\newcommand{\tr}{\mathrm{tr}}

\newcommand{\background}{(\mathring{j}_f,\mathring{g}_{ve},\mathring{z}_{vf})}
\newcommand{\sgn}{\mathrm{sgn}}

\begin{document}

\sloppy

\title{\bf Covariant Loop Quantum Gravity, Low Energy Perturbation Theory, and Einstein Gravity with High Curvature UV Corrections}

\author{Muxin Han }

\affiliation{Centre de Physique Th\'eorique
, CNRS UMR7332, Aix-Marseille Universit\'e and Universit\'e de Toulon, 13288 Marseille, France}


\begin{abstract}

A low-energy perturbation theory is developed from the nonperturbative framework of covariant Loop Quantum Gravity (LQG) by employing the background field method. The resulting perturbation theory is a 2-parameter expansion in the semiclassical and low-energy regime. The two expansion parameters are the large spin and small curvature. The leading order effective action coincides with the Einstein-Hilbert action. The subleading corrections organized by the two expansion parameters give the modifications of Einstein gravity in quantum and high-energy regime from LQG. The perturbation theory developed here shows for the first time that covariant LQG produces the high curvature corrections to Einstein gravity. This result means that LQG is not a naive quantization of Einstein gravity, but rather provides the UV modification. The result of the paper may be viewed as the first step toward understanding the UV completeness of LQG.

\end{abstract}


\maketitle


The nonperturbative covariant formulation of LQG adapts the idea of path integral quantization to the framework of LQG \cite{sfrev}. In the formulation, a \emph{spinfoam amplitude} $A(\ck)$ is defined on a given simplicial manifold $\ck$ for the transition of boundary quantum 3-geometries (spin-network states in LQG) \cite{footnote2}. The spinfoam amplitude sums over the history of spin-networks, and suggests a foam-like quantum spacetime structure. 

In this paper, a low-energy perturbation theory is developed from the nonperturbative framework of LQG. The perturbation theory explains how classical gravity emerges from the group-theoretic spinfoam formulation, and provides the high-energy (high curvature) and quantum corrections. Importantly the perturbation theory developed here shows for the first time that  covariant LQG produce the high curvature corrections, which modifies the UV behavior of Einstein gravity. And it is the first time that a systematic way is developed to compute the high curvature corrections from a full LQG framework. 

The discussion here focuses on the Lorentzian spinfoam amplitude proposed by Engle-Pereira-Rovelli-Livine (EPRL) \cite{EPRL}. The nonperturbative construction of EPRL spinfoam amplitude is purely (quantum-)group-theoretic. As one of the representations \cite{BHKKL}, the EPRL spinfoam amplitude reads
\be
A(\ck)=\sum_{J_f}d_{J_f}\tr\lt[\prod_eP^{\mathrm{inv}}_e\rt]
\ee
$P^{\mathrm{inv}}_e$ is a invariant projector onto a certain subspace of the $\Slc$ intertwiners, associated to each tetrahedron $e$ in $\ck$.  Here $f$ labels a triangle in $\ck$, $e$ labels a tetrahedron, and $v$ labels a 4-simplex. $J_f$ is the SU(2) spin assigned to each triangle. $d_J$ is the dimension of the SU(2) irrep with spin $J$. The above nonperturbative spinfoam amplitude is \emph{finite} in the quantum group version \cite{QSF}, which includes the cosmological constant in LQG \cite{CC}.

The EPRL spinfoam amplitude can be written in the following path-integral-like expression (see \cite{hanPI} for a derivation):
\be
A(\ck)&=&\sum_{J_{f}}d_{J_f} \int_{\Slc}
\rmd g_{ve} \int_{\mathbb{CP}^{1}} \rmd{z_{vf}}\ e^{S[J_f,g_{ve},z_{vf}]}\label{A}
\ee
where $g_{ve}$ is a $\Slc$ group variable associated with each dual half-edge.  $z_{vf}$ is a 2-component spinor. The spinfoam action $S$ given by
\be
S\lt[J_f,g_{ve},z_{vf}\rt]=\sum_{(e,f)}\Big\{J_f\cv_f\lt[g_{ve},z_{vf}\rt]+i\g J_f \ck_f\lt[g_{ve},z_{vf}\rt]\Big\}
\label{action}
\ee
where the short-hand notations $\cv_f$ and $\ck_f$ are defined by
\be
\cv_f&\equiv& \ln\lt[{\langle Z_{vef},Z_{v'ef}\rangle^{2}}{ \langle Z_{v'ef},Z_{v'ef}\rangle^{-1} \langle Z_{vef},Z_{vef}\rangle^{-1}}\rt]\nonumber\\
\ck_f&\equiv&\ln \lt[\langle Z_{vef},Z_{vef}\rangle\langle Z_{v'ef},Z_{v'ef}\rangle^{-1}\rt]\label{VK}
\ee
with $Z_{vef}=g_{ve}^\dagger z_{vf}$. $\g\in \R$ is the Barbero-Immirzi paramter.

Practically, we apply the background field method to Eq.\Ref{A} and consider the perturbations of the spinfoam variables around a given background configuration (spinfoam data $(J_f,g_{ve},z_{vf})$ on $\ck$)\cite{footnote1}. The perturbative expansion is performed in the semiclassical and low-energy regime. Such a regime can be specified in the following way: The existing semiclassical results suggest that the semiclassical geometry emerges from spinfoam is discrete with a (triangle-area) spacing $\a_f=\g J_f\ell_P^2$ \cite{semiclassical,CF,HZ,hanPI}. Here we focus on the regime that the area scale $\a_f$ is much greater than the Planck scale $\ell_P^2$, but much smaller than the mean curvature radius $L^2$ of the semiclassical geometry, i.e. 
\be
\ell_P^2\ll\a_f\ll L^2\label{regime}
\ee
Eq.\Ref{regime} is a 4d analog of the semiclassical regime in canonical LQG \cite{QCS}. The relation $\ell_P^2\ll\a_f$ comes from $\hbar\to0$ and implies the semiclassicality. $\a_f\ll L^2$ implies the low-energy approximation, since it requires that the mean wave-length of the gravitational fluctuation is much larger than the lattice scale. Adapting Eq.\Ref{regime} to the spinfoam formulation, $\ell_P^2\ll\a_f$ can be implemented by $J_f\gg1$ for all $f$, while $\a_f\ll L^2$ means that the deficit angle $|\Theta_f|\ll1$ for all $f$, because $|\Theta_f|\sim\a_f/L^2[1+o(\a_f/L^2)]$ \cite{FFLR}. In the following, the perturbative analysis of spinfoam amplitude $A(\ck)$ is performed with respect to a certain background spinfoam configuration in the semiclassical low-energy regime Eq.\Ref{regime}. The analysis results in a low-energy effective action, whose leading contribution coincides with the Einstein-Hilbert action. The expansion parameters $J_f^{-1}$ and $\Theta_f$ organize respectively the quantum and high-energy curvature corrections.


Let's consider the spinfoam amplitude in the regime $\ell_P^2\ll\a_f$. We write Eq.\Ref{A} by $A(\ck)=\sum_{J_f}d_{J_f} A_{J_f}(\ck)$ and focus on the summing over fluctuations of $J_f$ in the large-$J$ regime. The partial amplitude $A_{J_f}(\ck)$ has been defined by collecting the ($g,z$)-integrals in Eq.\Ref{A}. The spins $J_f\equiv\l j_f$ is large for all $f$, where $\l\gg1$ is the mean value of $J_f$. By the linearity of $S[J_f,g_{ve},z_{vf}]$ in $J_f$, the stationary phase analysis is employed to study the asymptotic behavior of the partial amplitude $A_{J_f}(\ck)$ as $J_f$ uniformly large. Such an analysis has been developed in \cite{semiclassical,CF,HZ,hanPI}. In the asymptotics, the leading contribution of $A_{J_f}(\ck)$ comes from the spinfoam critical configurations, i.e. the solutions of $\Re S=0$ and $\partial_g S=\partial_z S=0$. It turns out that each critical configuration is interpreted as a certain type of geometry on $\ck$. Moreover the critical configurations also know if the manifold is oriented and time-oriented \cite{hanPI}. As a result, the critical configurations are classified according to their geometrical interpretations and the information about orientations:
\begin{center}
\vspace{-0.3cm}
\begin{tabular}{| l || c | r |}
\hline
  & $\cv_f$ & $\ck_f$\ \ \ \ \ \ \ \\ \hline \hline
\ Lorentz Time-Oriented&\ \ 0& $\eps\ \sgn(V_4)\Theta_f$\  \\ \hline
Lorentz Time-Unoriented&\ \ $i\eps\pi$ & $\eps\ \sgn(V_4)\Theta_f$\  \\ \hline
\ \ \ \ \ \ \ \ \ \ \ Euclidean & $i\eps \lt[\sgn(V^E_4)\Theta^E_f+\pi n_f\rt]$ & 0\ \ \ \ \ \ \ \ \\ \hline
\ \ \ \ \ \ \ \ \ \ \ \ \ Vector & $i\Phi_f$ &  0\ \ \ \ \ \ \ \  \\ \hline
  \end{tabular}\\
  \vspace{0.4cm}
\vspace{-0.3cm}
\end{center}
The first 2 classes of critical configurations give the Lorentzian simplicial geometries on $\ck$. Each critical configuration $(j_f,g_{ve},z_{vf})$ in the first 2 classes is equivalent to a set of geometrical data $(\pm_v E_\ell(v),\eps)$ \cite{hanPI} with $\eps=\pm1$. $E_\ell(v)$ is a cotetrad on $\ck$ (the edge-vectors satisfying some conditions), up to a overall sign $\pm_v$ in each 4-simplex. $E_\ell(v)$ determines the oriented volume $V_4(v)=\det\lt(E^I_\ell(v)\rt)$. The local spacetime orientation is defined by $\sgn(V_4)$. $E_\ell(v)$ also determines uniquely a spin connection $\O_{e}\in \mathrm{SO}(1,3)$ along each dual edge $e$. The critical configuration gives a locally time-oriented spacetime if the cooresponding spin connection along a closed loop $\O_f=\prod_{e\subset\partial f}\O_e\in \mathrm{SO}^+(1,3)$. Additionally, the last 2 classes of critical configurations give the Euclidean simplicial geometry and degenerate vector geometry on $\ck$. It turns out that $\cv_f$ and $\ck_f$ defined in Eq.\Ref{VK} take different values in each class of critical configurations, as is shown in the above table. Here $\Theta_f$ ($\Theta_f^E$) denotes the Lorentzian (Euclidean) deficit angle, $\Phi_f$ denotes the vector-geometry angle, and $n_f\in\{0,1\}$.


Here we consider the perturbations of the spinfoam variables around a critical configuration $\background$ in the 1st class, which corresponds to the globally oriented and time-oriented Lorentzian simplicial geometry with $\sgn(\mathring{V}_4)=1,\mathring{\eps}=-1$ globally. It turns out that Einstein gravity is recovered from the perturbations around such a background. The background deficit angles $|\mathring{\Theta}_f|\ll1$ since we are interested in the low-energy perturbations. The background spins $\mathring{J}_f=\l \mathring{j}_f$ with $\l\gg1$, for the semiclassical approximation. 

The partial amplitude can be written as $A_{j_f}(\ck)=\exp \l W[j_f]$, where $W[j_f]$ is an effective action obtained by integrating out the $(g_{ve},z_{vf})$-variables in Eq.\Ref{A}. $W[j_f]$ is computed in a neighborhood at the background $\background$, by generalizing the method of computing effective action to the case of a complex action \cite{almostanalytic} (sometimes called almost-analytic machinery). 
\be
W[j_f]=\cs[j_f;g_{ve}(j),\tilde{g}_{ve}(j);z_{vf}(j),\tilde{z}_{vf}(j)]+\cdots
\ee
where $\cdots$ stands for the subleading contributions of $o(1/\l)$. $\cs[j_f;g_{ve},\tilde{g}_{ve};z_{vf},\tilde{z}_{vf}]$ is the analytic continuation of the action $S[j_f,g_{ve},z_{vf}]$ in a complex neighborhood at $\background$. The leading contribution of $W[j]$ is given by evaluating $\cs$ at the solution $\{g_{ve}(j),\tilde{g}_{ve}(j);z_{vf}(j),\tilde{z}_{vf}(j)\}\equiv Z(j)$ of $\partial_g\cs=\partial_{\tilde{g}}\cs=\partial_z\cs=\partial_{\tilde{z}}\cs=0$. 
In a neighborhood of spins at $\mathring{j}_f$, the real part of $\cs[j_f;Z(j)]$ is nonvanishing and negative unless $j_f=\mathring{j}_f$, where $Z(j)$ reduces to the real value $\mathring{g}_{ve},\mathring{z}_{vf}$.

The leading contribution $\cs[j_f;Z(j)]$ can be analyzed by Taylor expansion in perturbations $\fs_f=j_f-\mathring{j}_f$: 
\be
\cs=i\lt[\sum_f\g\mathring{j}_f\mathring{\Theta}_f+\sum_f\g\mathring{\Theta}_f\fs_f+\sum_{f,f'}W_{f,f'}\fs_f\fs_{f'}+o\lt(\fs^3\rt)\rt]\label{7}
\ee
The computations of the above coefficients at different orders are given in \cite{statesum}. $W_{f,f'}$ is local in the sense that it vanishes unless $f,f'$ belong to the same tetrahedron $e$.

The above result is for the partial amplitude $A_{j_f}(\ck)$. In order to compute $A(\ck)$, we implement the sum over perturbations $\fs_f$ inside a neighborhood at $\mathring{j}_f$. The spinfoam amplitude is written as $A(\ck)\sim\sum_{\fs_f}d_{\l(\mathring{j}_f+\fs_f)}\exp \l W[\mathring{j}_f+\fs_f]$ and is studied perturbatively. The Poisson resummation formula can be applied to the sum over the perturbations $\fs_f$, which results in the following perturbative expression for $A(\ck)$:
\be
e^{i\l\sum_f\g\mathring{j}_f\mathring{\Theta}_f}\sum_{k_f\in\Z}\int\lt[\rmd \fs_f\rt]  e^{i\l\lt[\sum_f\lt(\g\mathring{\Theta}_f-4\pi k_f\rt)\fs_f+\sum_{f,f'}W_{f,f'}\fs_f\fs_{f'}+o\lt(\fs^3\rt)\rt]+\cdots}\label{branches}
\ee 
where again $\cdots$ stands for the subleading contributions in $1/\l$. 

The above discussion considers the large-$J$ regime for the spinfoam amplitude for the semiclassical approximation. Now we implement the low-energy approximation. The low-energy regime is achieved when the background configuration $\background$ is such that $\mathring{\Theta}_f\ll1$.  

Firstly let's consider the integrals with $k_f\neq0$ in Eq.\Ref{branches} and apply the stationary phase analysis as $\l\gg1$. The equation of motion from $\cs[j_f;Z(j_f)]$ is given by
\be
0=\partial_{j_f}\cs|_{Z(j)}+\partial_{j_{f}}Z\partial_Z\cs|_{Z(j)}=\partial_{j_f}\cs|_{Z(j)}
\ee
where $\partial_Z\cs|_{Z(j)}=0$ because $Z(j)$ is the solution of $\partial_g\cs=\partial_{\tilde{g}}\cs=\partial_z\cs=\partial_{\tilde{z}}\cs=0$. The condition $\Re\cs[j_f;Z(j_f)]=0$ implies the perturbation $\fs_f=0$ where $Z(j)$ reduces to $\mathring{g}_{ve},\mathring{z}_{vf}$. Taking into account both the equations of motion and $\Re\cs=0$ results in that $\g\mathring{\Theta}_f-4\pi k_f=0$ for $k_f\neq0$, which cannot be satisfied in the low-energy regime where $|\mathring{\Theta}_f|\ll1$ (with $\g\sim o(1)$ or less). As a result, all the integrals with $k\neq0$ in Eq.\Ref{branches} are exponentially decaying, according to the principle of stationary phase analysis \cite{stationaryphase}. 

We thus focus on the integral with $k_f=0$ in Eq.\Ref{branches}:
\be
\int\lt[\rmd \fs_f\rt]  e^{i\l\lt[\sum_f\g\mathring{\Theta}_f\fs_f+\sum_{f,f'}W_{f,f'}\fs_f\fs_{f'}+o\lt(\fs^3\rt)\rt]+\cdots}.\label{1}
\ee
We denote by $|\mathring{\Theta}|\ll1$ the mean value of the background deficit angle and $\mathring{\Theta}_f=\mathring{\Theta}\Delta_f$. The 2d space of $(\l,\mathring{\Theta})$ may be viewed as the parameter space for our perturbation theory, where the semiclassical and low-energy regime is located in $\l\gg1,|\mathring{\Theta}|\ll1$. Now a new parameter is defined by $\b:=\l\mathring{\Theta}$, or a coordinate transformation is defined from $(\l,\mathring{\Theta})$ to $(\l,\b)$, where $\b$ is treated independent of $\l$. Then Eq.\Ref{1} reads
\be
\int\lt[\rmd \fs_f\rt]  e^{i\l\lt[\sum_{f,f'}W_{f,f'}\fs_f\fs_{f'}+o\lt(\fs^3\rt)\rt]}e^{i\b\g\sum_f\Delta_f\fs_f+\cdots}\label{2}
\ee
Again the stationary phase analysis is applied as $\l\gg1$. We find $\fs_f=0$ is a solution of both $\partial_{\fs_f}[\sum_{f,f'}W_{f,f'}\fs_f\fs_{f'}+o(\fs^3)]=0$ and $\Re[\sum_{f,f'}W_{f,f'}\fs_f\fs_{f'}+o(\fs^3)]=0$. Note that $\Re[\sum_{f,f'}W_{f,f'}\fs_f\fs_{f'}+o(\fs^3)]=\Re\cs$ since $i\sum_f\g\mathring{\Theta}_f\fs_f$ is purely imaginary. The standard stationary phase formula \cite{stationaryphase} leads to the following result from Eq.\Ref{2} in the neighborhood of the background spins $\mathring{j}_f$ ($\fs_f=0$): 
\be
\sum_{n=0}^\infty\lt(1/\l\rt)^n\Fl_n\lt[e^{i\b\g\sum_f\Delta_f\fs_f+\cdots}\rt]_{\fs_f=0}
=\sum_{n=0}^\infty\sum_{r=0}^{2n}\lt(\g^r\b^r/\l^n\rt)f_{n,r}\label{powercounting}
\ee
$\Fl_n$ is a differential operator of order $2n$ (in $\partial_{\fs_f}$) where all the interactions from the Lagrangian are encoded (see \cite{stationaryphase} for a general expression). Applying the differential operator $\Fl_n$ to $e^{i\b\g\sum_f\Delta_f\fs_f}$ gives the power-counting result in Eq.\Ref{powercounting}. The coefficients $f_{n,r}$ are functions of $\l$ and $\background$, which are regular as $\l\to\infty$ \cite{footnote0}. Inserting Eq.\Ref{powercounting} to Eq.\Ref{branches} and recalling $\b=\l\mathring{\Theta}$, the following expansion for $A(\ck)$ is obtained:
\be
A(\ck)\sim e^{i\l\sum_f\g\mathring{j}_f\mathring{\Theta}_f}\sum_{n=0}^\infty\sum_{r=0}^{2n}\lt(\g^r\mathring{\Theta}^r/\l^{n-r}\rt)f_{n,r}
\ee
where the exponentially decaying contributions have been neglected. We can read from the above result an effective action $I_{\mathrm{eff}}\background$ by expressing $A(\ck)\sim \exp iI_{\mathrm{eff}}$, where the effective action at the background $\background$ is an expansion w.r.t. $\mathring{\Theta}$ and $\l^{-1}$: 
\be
iI_{\mathrm{eff}}=\l\lt[i\sum_f\g\mathring{j}_f\mathring{\Theta}_f+\frac{\g^2}{4}\sum_{f,f'}W^{-1}_{f,f'}\mathring{\Theta}_f\mathring{\Theta}_{f'}+o(\g^3\mathring{\Theta}^3;\l^{-1})\rt]\label{EA}
\ee
The coefficient $W^{-1}_{f,f'}$ is the inverse of $W_{f,f'}$ in Eq.\Ref{7}. $W_{f,f'}$ is nonzero only when $f,f'$ belong to the same tetrahedron $e$
\be
W_{f,f'}=\frac{2(1+2i\g-4\g^2-2i\g^3)}{5+2i\g}\hat{n}_{ef}^t\mathbf{X}^{-1}_e\hat{n}_{ef'}
\ee
where $\mathbf{X}_e^{ij}\equiv\sum_fj_f(-\delta^{ij}+\hat{n}^i_{ef}\hat{n}_{ef}^j+i\eps^{ijk}\hat{n}_{ef}^k)$. Here the unit 3-vector $\hat{n}_{ef}$ determined by $\background$ is the normal vector of the triange $f$ in the frame of the tetrahedron $e$ \cite{statesum,hanPI}. Although $W_{f,f'}$ is local in $f,f'$, the inverse $W_{f,f'}^{-1}$ is nonlocal in general, it may be nonzero for far away $f,f'$. So the $\g^2\mathring{\Theta}^2$-term is a nonlocal curvature correction in $I_{\mathrm{eff}}$. Moreover there is a systematic way developed in \cite{statesum} to compute in principle all the $\g^r\mathring{\Theta}^r$-corrections.

There are several remarks for the effective action Eq.\Ref{EA}:

\noindent
\textbf{Low-energy effective action as curvature expansion:} 
The terms $\propto\l(\g\mathring{\Theta})^{r\geq2}$ are understood as the high-energy correction to the leading order $i\l\sum_f\g\mathring{j}_f\mathring{\Theta}_f$, since the $|\mathring{\Theta}|\ll1$ implements the low-energy approximation \cite{footnote4}. Therefore as a power-series of $\mathring{\Theta}$, $I_{\mathrm{eff}}$ is understood as a low-energy effective action from covariant LQG. The deficit angle $\mathring{\Theta}\sim\a \calr$, where $\a$ is the mean (area-)spacing of the lattice given by the background data $\background$, $\calr$ is the mean curvature of the background. Thus the effective action $I_{\mathrm{eff}}$ can be viewed as a curvature expansion, where the high-energy corrections are given by $\a^2\g^2\calr^2+\a^3\g^3\calr^3+\cdots$ with $\a$ being the (effective) coupling constant of the high-derivative interactions.

\noindent
\textbf{2-parameter expansion:} There are two parameters involved in the expression of effective action $I_{\mathrm{eff}}$, i.e. $\l\gg1$ and $\mathring{\Theta}\ll1$ (or $\a$ with dimension $-2$). $1/\l$ counts the quantum corrections, while $\mathring{\Theta}$ (or $\a$) counts the high-energy corrections. The two expansion parameters implements the semiclassical low-energy regime $\ell_P^2\ll\a\ll L^2$. 

\begin{figure}[h]
\begin{center}
\includegraphics[width=5cm]{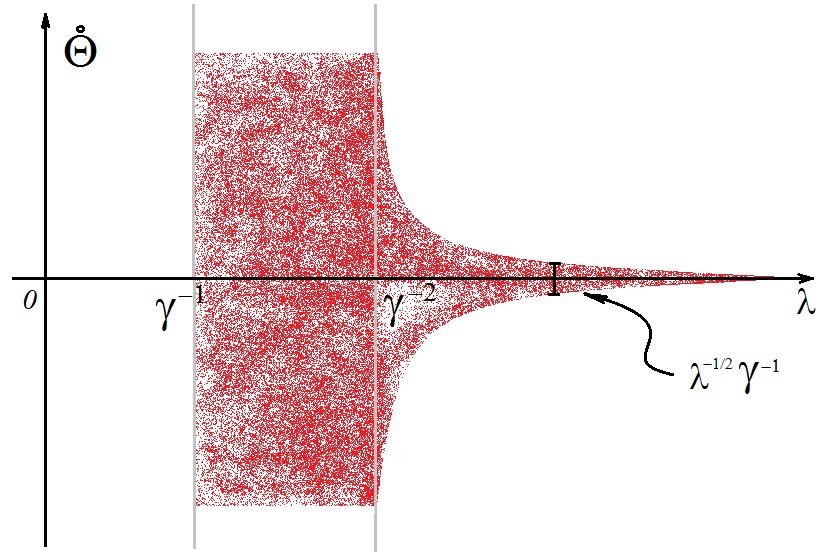}
\caption{The Einstein sector of spinfoam configurations.}
  \label{phases1}
  \end{center}
\end{figure}

\noindent
\textbf{Restriction of $\mathring{\Theta}$:} The effective action $iI_{\mathrm{eff}}$ has a negative real part, which is contained in the terms of higher-curvature \cite{statesum}, i.e. $\Re[iI_{\mathrm{eff}}]=\l\Re[\frac{1}{4}W^{-1}\g^2\mathring{\Theta}^2+o(\g^3\mathring{\Theta}^3)+\cdots]\leq0$ where $\cdots$ stands for the terms suppressed by $1/\l$. This negative real part on the exponential would have given an exponentially decaying factor in $A(\ck)$ if $\g\mathring{\Theta}$ was of $o(1)$, which is \emph{not} our case because of $\mathring{\Theta}\ll1$. The non-decaying $A(\ck)$ requires that $\Re[iI_{\mathrm{eff}}]$ doesn't scale to be large by $\l\gg1$, which results in a nontrivial bound of the deficit angle $\mathring{\Theta}$, i.e.
\be
|\mathring{\Theta}|\leq \g^{-1}\l^{-\half},\label{bound}  
\ee
The situation is illustrated in FIG.\ref{phases1}. The red region in FIG.\ref{phases1} illustrates the space (in the coordinates $\l$ and $\mathring{\Theta}$) of background configurations $(\mathring{J}_f,\mathring{g}_{ve},\mathring{z}_{vf})$, which validates the 2-parameter expansion of the effective action $I_{\mathrm{eff}}$. If $\mathring{\Theta}$ is beyond the bound Eq.\Ref{bound}, where the approximation Eq.\Ref{powercounting} is invalid, the integral Eq.\ref{1} is exponentially decaying as $\l\gg1$ by the same argument for $k\neq0$ integrals. Thus the red region in FIG.\ref{phases1} illustrates the semiclassical low-energy effective degrees of freedom from the above approximation.

\noindent
\textbf{Einstein-Hilbert action:} After the restriction Eq.\Ref{bound}, the leading contribution in $I_{\mathrm{eff}}$, $i\l\sum_f\g\mathring{j}_f\mathring{\Theta}_f$, is the Regge action of GR as a functional of the edge-lengths determined by $\background$ (by identifying $\g\l\mathring{j}_f=\a_f/\ell_P^2$ to be the area of the triangle $f$ in Planck unit). Moreover given that that $\mathring{\Theta}_f\sim\a_f/L^2\ll1$ \cite{footnote} and that $I_{\mathrm{eff}}$ is a power-series in $\a_f$, the leading order contribution is essentially the Einstein-Hilbert action on a smooth manifold $\cm$, i.e. as a functional (see e.g. \cite{FFLR})
\be
i\l\sum_f\g\mathring{j}_f\mathring{\Theta}_f=\frac{i}{\ell_P^2}\int_{\cm}\rmd^4x\sqrt{-\mathring{g}}\calr[\mathring{g}_{\a\b}]\times\lt[1+o\lt(\a_f/L^2\rt)\rt]
\ee 
where $\mathring{g}_{\a\b}$ is the Lorentzian metric approximated by the simplicial geometry from $\background$. Therefore $I_{\mathrm{eff}}$ can be written as
\be
I_{\mathrm{eff}}=\frac{1}{\ell_P^2}\int_{\cm}\rmd^4x\sqrt{-\mathring{g}}\calr[\mathring{g}_{\a\b}]\times\lt[1+o\lt(\a\calr\rt)+o(1/\l)\rt]\label{20}
\ee 
where the leading contribution is Einstein-Hilbert action. The diffeomorphism invariance on $\cm$ is then recovered as an approximated symmetry in the leading order of semiclassical and low energy approximation.

\noindent
\textbf{Small Barbero-Immirzi parameter:} Once $\g\ll1$, an interesting regime $\g^{-1}\ll\l\leq\g^{-2}$ appears in FIG.\ref{phases1}. $\g^{-1}\ll\l$ is required for $\ell_P^2\ll\a_f$. As $\l\ll\g^{-2}$ and Eq.\Ref{bound}, $A(\ck)$ is not decaying even without the restriction of $\mathring{\Theta}$. Even a finite $\mathring{\Theta}$ is admitted in Eq.\Ref{EA} without requiring $\mathring{\Theta}\ll1$. Indeed each $\mathring{\Theta}$ is accompanied by a $\g$ in $I_{\mathrm{eff}}$ (and in Eq.\Ref{1} originally), where $\g$ appears as an effective scaling of the deficit angle. One may choose $\b=\l\g$ in Eq.\Ref{2} as $\g\ll1$. Thus in the regime $\g^{-1}\ll\l\leq\g^{-2}$, $I_{\mathrm{eff}}$ can be formulated as Eq.\Ref{EA} with finite deficit angle, where the leading order is Regge action in general. Sending $\g\to0$ neglects effectively the higher-curvature corrections. The analysis here may explain the spinfoam graviton propagator calculations \cite{propagator1,propagator2,3pt} and the analysis in \cite{claudio}, which firstly motivate $\g\ll1$.

\noindent
\textbf{Flatness:} $\l\to\infty$ asymptotically is another interesting regime in FIG.\Ref{phases1}, where the deficit angle is so restricted that only $\mathring{\Theta}=0$ (flat geometry) is allowed. It relates to the ``flatness problem'' in spinfoam formulation discussed in \cite{flatness}. However the flatness problem disappears here for any finite $\l\gg1$ by the low-energy perturbation theory \cite{footnote5}. 

The above discussion considers the fluctuations of spinfoam variables which touches a single critical configuration $\background$. The fluctuations touching many critical configurations $(j_f,g_{ve},z_{vf})_c$ results in a sum of the above perturbative expression of $A(\ck)$ over all the critical configurations. We name the red region in FIG.\ref{phases1} the \emph{Einstein sector} $\Fn_E$ as a subspace of spinfoam configurations, in which all the critical configurations are interpreted as globally oriented and time-oriented Lorentzian geometry (with $\sgn(V_4)=1,\eps=-1$ globally). When the fluctuations of the spinfoam variables are considered within $\Fn_E$, the perturbative expression of the spinfoam amplitude is then given by
\be
A(\ck)=\sum_{(j_f,g_{ve},z_{vf})_c\in \Fn_E}e^{\frac{i}{\ell_P^2}\int_{\cm}\rmd^4x\sqrt{-{g}}\calr[{g}_{\a\b}]\times\lt[1+o\lt(\a\calr\rt)+o(1/\l)\rt]} \label{sum}
\ee
where ${g}_{\a\b}$ is the Lorentzian metric approximated by $(j_f,g_{ve},z_{vf})_c$. Eq.\Ref{sum} makes sense because the perturbations at a geometrical critical configuration (of the type globally Lorentzian, oriented, and time-oriented) only touch the geometrical critical configuration of the same type. From Eq.\Ref{sum} we see that the contributions to $A(\ck)$ from the perturbations within $\Fn_E$ are given by the functional integration of Einstein-Hilbert action (with a discrete measure) plus the high-energy and quantum corrections. The leading contributions to $A(\ck)$ in $\Fn_E$ come from the critical configurations $(j_f,g_{ve},z_{vf})_c$ which give ${g}_{\a\b}$ satisfying Einstein equation (with high-energy and quantum corrections).

The above discussion can be generalized straight-forwardly to the analysis of correlation functions. In the Einstein sector $\Fn_E$, the perturbative result of the spinfoam correlation function coincides with the corresponding perturbative correlation function from Einstein gravity or Regge gravity, up to curvature and quantum corrections.

The corrections of higher order in curvature (in deficit angle) modifies the Einstein (Regge) gravity in high-energy regime. It is interesting to further investigate these high-curvature terms predicted from covariant LQG, in order to see if LQG can provide a UV-completion of perturbative Einstein gravity. The origin of high-curvature terms is the sum over \emph{non-Regge-like} spins (the spins that cannot be viewed as Regge areas) in spinfoam amplitude. The non-Regge-like spins are the extra UV degrees of freedom in addition to GR predicted by LQG. Their dynamics may be studied via the action Eq.\Ref{7} to see if they regulate Einstein gravity at UV.

Finally we remark that the analysis beyond the Einstein sector $\Fn_E$ can also be carried out. There exists different other sectors, well-separated from $\Fn_E$, where the similar analysis results in the leading order effective actions different from Einstein gravity. We refer to \cite{statesumgamma,statesum} for detailed discussions. 


\section*{Acknowledgments}
 
The author thanks C. Rovelli for proof-reading and comments. He also thanks the anonymous referees for their comments, which helps the author to improve the presentation. The research has received funding from the People Programme (Marie Curie Actions) of the European Union's 7th Framework Programme (FP7/2007-2013) under REA grant agreement No. 298786.


\begin{thebibliography}{20}
\bibitem{footnote2}The spinfoam amplitude $A$ is a $\ch$-valued function on the space of simplicial manifolds, where $\ch$ is the boundary Hilbert space and $\ch=\C$ if the manifold has no boundary.
\bibitem{footnote0}If the $\l^{-1}$-corrections are neglected, the $\g\mathring{\Theta}$-expansion of $I_{\mathrm{eff}}$ is analytic in a neighborhood at $\g\mathring{\Theta}=0$, by the analyticity of the spinfoam action \cite{statesum}. 
\bibitem{footnote4}The terms linear to $\g\mathring{\Theta}$ are suppressed by $\l^{-1}$ except the leading Regge action.
\bibitem{footnote5}It is an open question about the interpretation for the regime of too large spins, which seems to give a too large lattice spacing scale $\a_f\sim\l\ell_P^2$ semiclassically, and contradict the observation of smooth spacetime. In order to remove the regime, a spin cut-off may be introduced via q-deformation \cite{QSF}, which produce a relatively large bare cosmological constant \cite{CC}.
\bibitem{footnote}Given $\background$ with nontrivial mean curvature radius, in order to obtain Eq.(\ref{20}), a large triangulation is needed. e.g. If the size of $\ck$ measured by $\background$ is of the same order as the curvature radius $L$, the number of simplices is at least of the order $N\sim L^4/\alpha^2\gg1$.
\end{thebibliography}
\end{document}